\documentclass[aps,prd,twocolumn,floatfix,nofootinbib,superscriptaddress]{revtex4}

\usepackage{graphicx}
\usepackage{dcolumn}
\usepackage{bm}
\usepackage{amsmath}
\usepackage{graphicx}
\usepackage{enumerate}
\usepackage{appendix}
\usepackage{mathrsfs}
\usepackage{multirow}

\begin{document}

\title{Lensed fast radio bursts as a probe of time-varying gravitational potential induced by wave dark matter}

\author{Ran Gao}
\affiliation{Institute for Frontiers in Astronomy and Astrophysics, Beijing Normal University, Beijing 102206, China;}
\affiliation{School of Physics and Astronomy, Beijing Normal University, Beijing 100875, China;}

\author{Shuxun Tian}
\affiliation{School of Physics and Astronomy, Beijing Normal University, Beijing 100875, China;}

\author{Zhengxiang Li}
\email[]{zxli918@bnu.edu.cn}
\affiliation{Institute for Frontiers in Astronomy and Astrophysics, Beijing Normal University, Beijing 102206, China;}
\affiliation{School of Physics and Astronomy, Beijing Normal University, Beijing 100875, China;}

\author{He Gao}
\email[]{gaohe@bnu.edu.cn}
\affiliation{Purple Mountain Observatory, Chinese Academy of Sciences, Nanjing, 210023, People's Republic of China;}
\affiliation{Institute for Frontiers in Astronomy and Astrophysics, Beijing Normal University, Beijing 102206, China;}
\affiliation{School of Physics and Astronomy, Beijing Normal University, Beijing 100875, China;}

\author{Kai Liao}
\affiliation{Department of Astronomy, School of Physics and Technology, Wuhan university, Wuhan 430072, China;}

\author{Bing Zhang}
\affiliation{Nevada Center for Astrophysics, University of Nevada, Las Vegas, NV 89154, USA;}
\affiliation{Department of Physics and Astronomy, University of Nevada, Las Vegas, NV 89154, USA;}

\author{Zong-Hong Zhu}
\affiliation{Institute for Frontiers in Astronomy and Astrophysics, Beijing Normal University, Beijing 102206, China;}
\affiliation{School of Physics and Astronomy, Beijing Normal University, Beijing 100875, China;}
\affiliation{Department of Astronomy, School of Physics and Technology, Wuhan university, Wuhan 430072, China;}

\date{\today}

\begin{abstract}
Ultralight bosonic wave dark matter (DM) is preponderantly contesting the conventional cold DM paradigm in predicting diverse and rich phenomena on small scales. For a DM halo made of ultralight bosons, the wave interference naturally induces slow de Broglie time-scale fluctuations of the gravitational potential. In this paper, we first derive an estimation for the effect of a time-varying gravitational potential on photon propagation. Our numerical simulations suggest that the time-varying potential of a $10^{11}M_{\odot}$ halo composed of $10^{-22}\,\mathrm{eV}$ bosons would stretch or compress a time series signal by a factor of $10^{-10}$. Here, we propose that, due to the precise measurements of their arrival times, lensed repeating fast radio bursts (FRBs) have the potential to effectively validate temporal variations in gravitational potential by monitoring their images over a period of approximately $\mathcal{O}(1)$ years. With rapidly growing FRB observations, this method would serve as a promising method to directly probe the wave nature of galactic DM halos.
\end{abstract}
\maketitle

\section{Introduction} \label{sec:int}
Since dark matter (DM) was first proposed to interpret the measurements of rotation curves of galaxies~\citep{1933AcHPh...6..110Z,2009GReGr..41..207Z}, an overwhelming amount of observational data provides rich and compelling evidence for the presence of this mysterious component on a wide range of scales from (sub)galactic scales~\citep{1970ApJ...159..379R} to galaxy clusters~\citep{2006ApJ...648L.109C}, going up to large-scale structures~\citep{2013ApJS..208...20B}. However, the basic properties and identities of DM remain shrouded in mystery. Cold dark matter (CDM) used to be a promising paradigm with the preferred candidates being weakly interacting massive particles (WIMPs)~\citep{1985NuPhB.253..375S}. Over the past several decades, there has been a great experimental effort to constrain the properties of WIMP DM with the parameter space being very restricted~\citep{2018PhRvL.121k1302A,2020PhRvL.125q1802B,2021PhRvL.127z1802M,2023PhRvL.131d1002A}. Unfortunately, the interpretation of those constraints to the exclusion of WIMP models is not straightforward. Moreover, this beautiful and simple scenario is severely challenged by observations and simulations on small non-linear and galactic scales. These challenges also have been around for decades, such as the missing satellite problem~\citep{1999ApJ...516..530K,1999ApJ...522...82K}, the cusp-core problem~\citep{2010AdAst2010E...5D}, and the too-big-to-fail problem~\citep{2011MNRAS.415L..40B,2012MNRAS.422.1203B}.

Another class of DM models with rich phenomena on small scales has emerged as an appealing and popular class of candidates. These are ultra-light dark matter (ULDM) models, in which DM is composed by ultra-light particles with masses in the range $10^{-24}~{\mathrm{eV}}<m<\mathcal{O}(1)~\mathrm{eV}$. In this paradigm, a specific candidate is the QCD axion, which was introduced to address the strong CP problem of quantum chromodynamics and couples weakly to the standard model~\citep{1977PhRvL..38.1440P,1978PhRvL..40..223W,1978PhRvL..40..279W}. This prediction has inspired a great deal of experimental effort for its direct detection~\citep{2015ARNPS..65..485G,2021RvMP...93a5004S}. On (sub)galactic scales, such small masses of these DM particles form a condensate or a superfluid and exhibits wave phenomena. The ULDM (also usually referred as wave DM, fuzzy DM, or axion DM) models have been proposed as a promising solution to the above-mentioned problems on small scales. Meanwhile, wave DM behaves as CDM and maintains the observational successes of CDM on large scales. As a result, wave DM has gained increasing interest and attention as a viable candidate to account for the DM content of the universe~\citep{2021ARA&A..59..247H}.

Due to merits in predicting rich and intriguing phenomena, a large number of astrophysical observations have been proposed to probe the nature of wave DM and constrain the mass of ultralight bosonic particles, such as cosmic microwave/infrared background observations~\citep{2015PhRvD..91j3512H,2018MNRAS.476.3063H,2022PhRvD.105f3513F}, linear power spectrum and early structure formation~\citep{2015MNRAS.450..209B,2022MNRAS.510.1425K,2022MNRAS.515.5646D,2022JCAP...01..049L,2024ApJ...976...40W, 2017PhRvD..96l3514K,2021PhRvL.126g1302R}, galactic dynamics and structure~\citep{2019PhRvL.123e1103M,2020MNRAS.492..877E,2021PhRvD.103j3019C,2022PhRvD.106g5007W,2022PhRvD.106f3517D,2022PhRvD.106f3010G}, compact objects~\citep{2014JCAP...02..019K,2017NatPh..13..584C,2017PhRvL.119b1101F,2018PhRvD..98j2002P,2019PhRvL.123b1102D,2019PhRvD.100f3528E,2024A&A...685A..94E}, and photon propagation in axion background~\citep{2019PhRvL.122s1101F,2015JCAP...02..006P,2023PhRvL.130l1401L,2019JCAP...02..059I,2021PhRvL.126s1102B,2024PhRvD.109b1303G,2019PhRvD.100a5040F}. Among these current constraints, the
stellar cluster at the center of the ultrafaint dwarf Eridanus II was used to yield a limit of $m > 2 \times 10^{-20}\ \mathrm{eV}$~\citep{2019PhRvL.123e1103M}. However, it was argued that this constraint is sensitive to the assumption of the impact on the stellar cluster from the heating by de Broglie granules~\citep{2020MNRAS.492..877E,2021PhRvD.103j3019C}. In general, almost all these currently available constraints on $m$ are dependent on certain assumptions or sensitive to different sources of systematic uncertainties~\citep{2021ARA&A..59..247H}. In this context, the mass of the ultralight bosonic particles is still an open issue. Therefore, any new observational probes revealing the wave nature and constraining the mass of the boson are of important need.

For wave DM, there are two timescales of great interest: One is the fast Compton time scale of scalar field oscillation ~\citep{2014JCAP...02..019K}, and the other is the slow de Broglie timescale fluctuations due to wave interference (see the movies given by~\cite{PhysRevLett.121.151301}). Both these two effects lead to a time-varying gravitational potential. The first one can be searched for in pulsar timing array data~\citep{2014JCAP...02..019K,2018PhRvD..98j2002P,2020JCAP...09..036K}. However, methods for directly detecting the second de Broglie timescale variation of the gravitational potential induced by the wave interference of DM are still almost absent. In this paper, we propose to use strong lensing of repeating fast radio bursts (FRBs) as a powerful tool to probe this effect.

\section{Theoretical analysis} \label{sec:met}
This section presents the theoretical framework for the effect of a time-varying gravitational potential in a lensing system on photon propagation. For a static lensing system, two main characteristics are light deflection and magnification. The time-varying property of the lens is unimportant for traditional lensing analysis. However, as more and more burst signals with short durations have been reported, it is necessary to analyze the lensing properties for the time-varying potential. At present, such a general theoretical framework has not yet been established. A special case is the moving lens effect \citep{Birkinshaw1983.Nature.302.315}, in which the motion of lens induces a time-varying potential and results in a shift of photon frequency. Analogue to this effect, a time-varying potential would stretch a time series signal. Our aim is to achieve a formula to estimate the effect of a general time-varying potential on the stretching of a time series signal. 

Lensing with a time-varying potential is a three-dimensional issue. The first approximation we use is to describe this scenario in terms of a one-dimensional system. This means that we ignore the light deflection and the magnification associated with it. Here we just focus on the signal stretching, which can be manifested by the change of the frequency or energy of a single photon. 

In general relativity, the weak galactic gravitational field can be described by
\begin{equation}
    \mathrm{d}s^2 = -c^2(1+2\Phi/c^2)\mathrm{d}t^2 + (1-2\Phi/c^2)\mathrm{d}\mathbf{r}^2,
\end{equation}
where $\Phi=\Phi(x,t)$ is the Newtonian gravitational potential and $\Phi/c^2\ll1$. We consider a photon propagating along the $x$-axis. The geodesic equation can be directly calculated from the above metric. Keeping the linear term of $\mathcal{O}(\Phi/c^2)$, the time component reads
\begin{equation} \label{eq:02}
    \frac{\mathrm{d}^2t}{\mathrm{d}\tau^2} + \frac{1}{c^4} \frac{\partial\Phi}{\partial t}\bigg[c^2\bigg(\frac{\mathrm{d}t}{\mathrm{d}\tau}\bigg)^2 - \bigg(\frac{\mathrm{d}x}{\mathrm{d}\tau}\bigg)^2\bigg] + \frac{2}{c^2}\frac{\partial\Phi}{\partial x}\frac{\mathrm{d}t}{\mathrm{d}\tau}\frac{\mathrm{d}x}{\mathrm{d}\tau} = 0,
\end{equation}
where $\tau$ is an affine parameter that describes the motion of the photon. For the second term, the quantity in the parenthesis is zero on the background. This means the nonzero leading term in parenthesis is at the order of $\mathcal{O}(\Phi/c^2)$, and thus the total second term is proportional to $\mathcal{O}(\Phi^2/c^4)$ and can be ignored. For the remaining terms, we write $\mathrm{d}t/\mathrm{d}\tau$ as $E$, which denotes the photon energy. Considering the background motion, we can further set $\mathrm{d}x/\mathrm{d}\tau=1$, then Eq. (\ref{eq:02}) gives
\begin{equation}\label{eq:03}
    \frac{\mathrm{d}E}{\mathrm{d}x} + \frac{2E}{c^2}\frac{\partial\Phi}{\partial x} = 0.
\end{equation}
As we only consider the linear terms, we can regard $E$ as a constant in the second term in Eq. (\ref{eq:03}). Integrating this equation from the source (negative infinity) to observer (positive infinity) along the photon trajectory yields
\begin{equation}\label{eq:FirstIntegral}
    \frac{\Delta E}{E}=-\frac{2}{c^2}\int \frac{\partial\Phi}{\partial x} \mathrm{d}x,
\end{equation}
where $\Delta E$ is the energy difference between the source and observer. 

Physically, the lens potential tends to approach zero at infinity. If the lens potential is static, then $\partial/\partial x$ becomes $\mathrm{d}/\mathrm{d}x$ and Eq.~(\ref{eq:FirstIntegral}) gives zero. If the potential is time-dependent, there should leave a nonzero leading term. However, the static discussion indicates that the leading term is not clearly manifested in the $\int\partial\Phi/\partial x\,\mathrm{d}x$ term. Numerically, if we do the calculation directly using Eq.~(\ref{eq:FirstIntegral}), there will be a scenario where the subtraction of two large numbers yields a nonzero small value (due to catastrophic cancellation). To reveal the leading term, here we introduce an auxiliary equation $\int\mathrm{d}\Phi/\mathrm{d}x\,\mathrm{d}x=0$, where the integral is along the photon trajectory and from the source to observer. This is valid even if $\Phi$ is time-dependent. The difference between $\mathrm{d}\Phi/\mathrm{d}x$ and $\partial\Phi/\partial x$ is the $\partial\Phi/\partial t$ term. Therefore, in the lens conventions, the leading term of the integration of Eq. (\ref{eq:03}) gives the relative frequency shift
\begin{equation} \label{eq:dnu}
    z_i \equiv \frac{\Delta\nu_i}{\nu} = \mathcal{O}(1) \times \frac{1}{c^2}\int_{\mathrm{path},\ i}\frac{\partial\Phi}{\partial t}\mathrm{d}t,
\end{equation}
where we rewrite $E$ as the frequency $\nu$, $\Delta\nu_i$ is the frequency change of $i$-th image, the integration is along the $i$-th path, and the first term $\mathcal{O}(1)$ denotes a coefficient of order unity. The lens convention ($i$-th image) is quoted here for subsequent use. The estimation for the order of magnitude from Eq. (\ref{eq:dnu}) should be reasonable although the full three-dimensional analysis may slightly change the coefficient due to the bending of light. This is why we retain an unknown coefficient in Eq.~(\ref{eq:dnu}). A complete theoretical framework remains to be established in the future.

Equation (\ref{eq:dnu}) is consistent with the result about moving lens effect \citep{Birkinshaw1983.Nature.302.315}: $\Phi$ corresponds to the deflection angle and $\partial/\partial t$ corresponds to the moving speed of the lens. Shapiro time delay \citep{PhysRevLett.13.789} may also be seen as a clue supporting our result: Inserting $\partial/\partial t$ into the Shapiro delay gives our result. 

\begin{figure}
    \centering
    \includegraphics[width=0.95\linewidth]{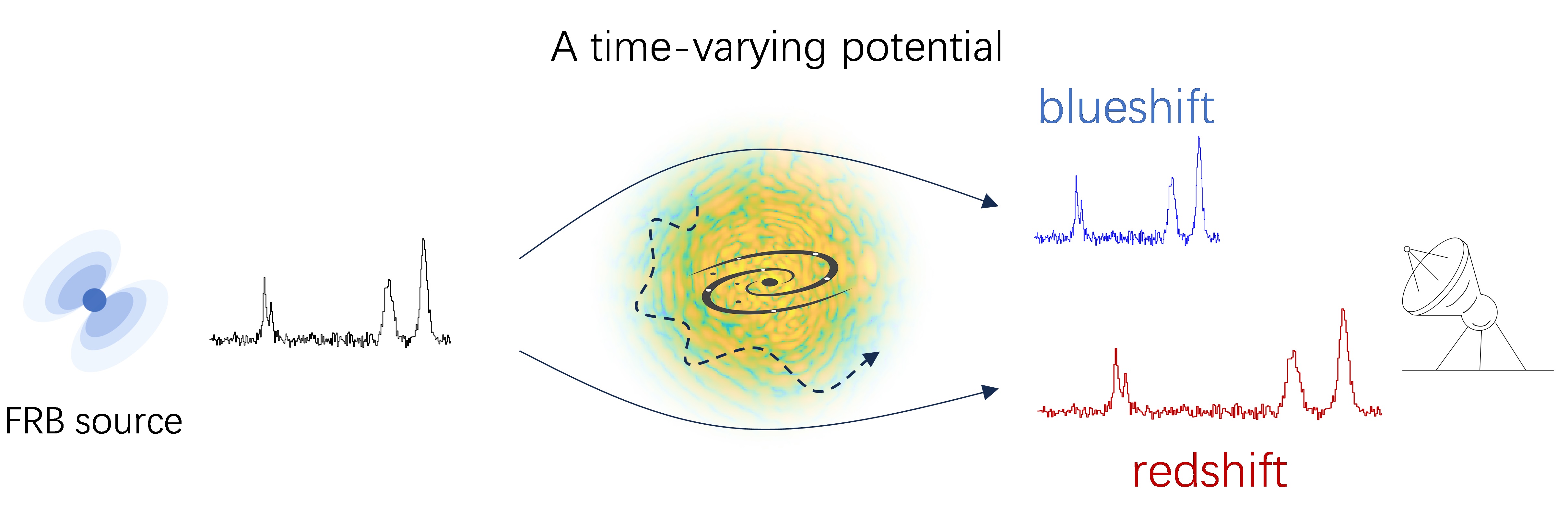}
    \caption{Illustrations for the effect of a time-varying gravitational potential induced by a wave DM halo and the scheme probing this effect with lensed repeating FRBs.}
    \label{fig:effect}
\end{figure}

It suggests that a time-varying gravitational potential causes a change in the frequency of the signals. Illustrations for the effect and our proposal for probing it are shown in Fig. \ref{fig:effect}, where sequence-signals (a series of bursts) of the two images experience stretching effects in different degrees as photons pass through a wave DM halo.
To quantify the magnitude of this effect, we carry out numerical simulations to obtain the magnitude of the de Broglie timescale change of the gravitational potential in a galactic DM halo composed of ultralight bosons.

\section{Simulations and estimations} \label{sec:sim}
\begin{figure}
    \centering
    \includegraphics[width=0.95\linewidth]{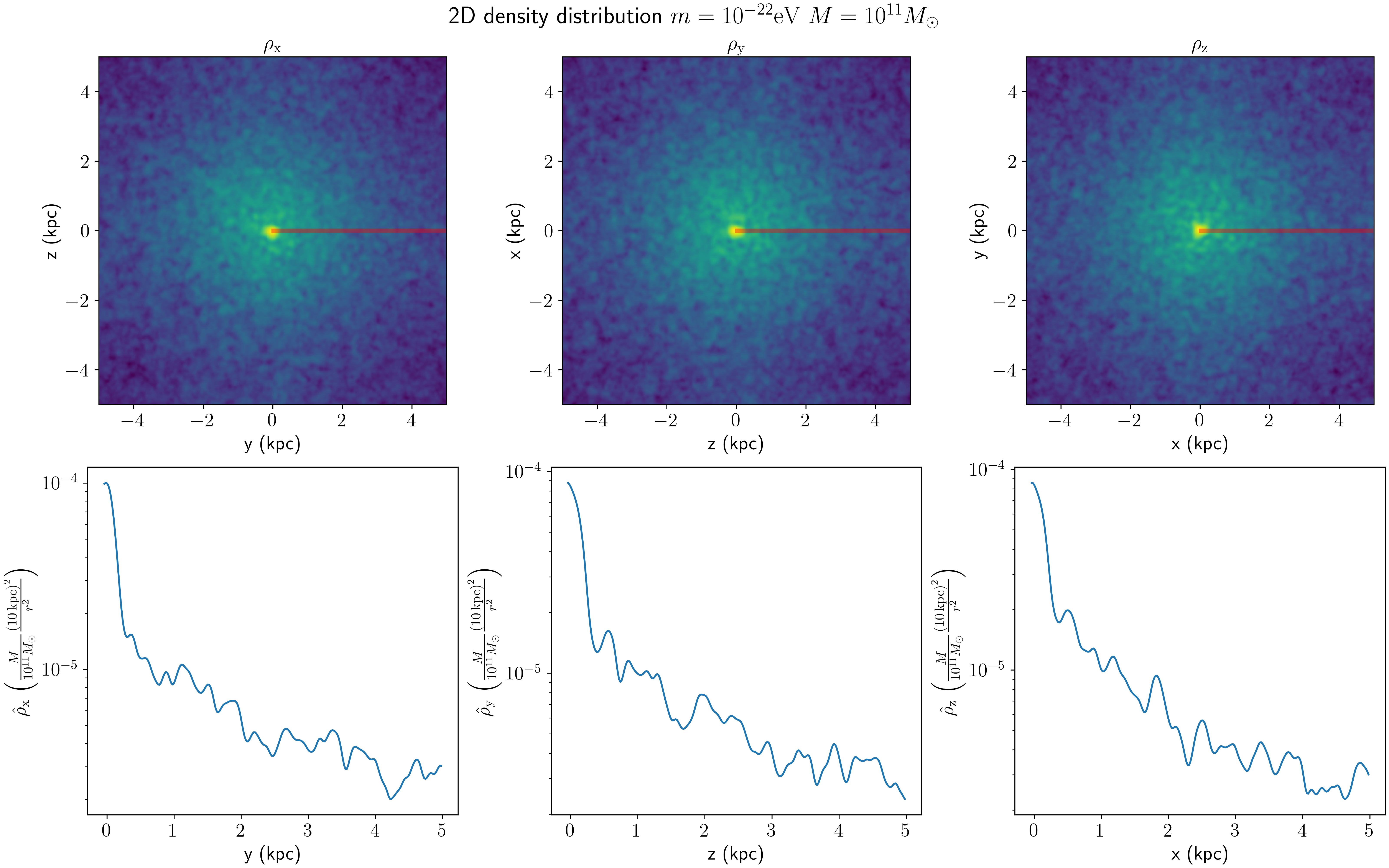}
    \caption{The 2D density distribution at the end of the stage 1 simulation. The top 3 plots show the 2D density distribution summed along the 3 axes, and the bottom 3 plots show the normalised density values in the shaded areas of the top plots, averaged along the longitudinal direction.}
    \label{fig:M11}
\end{figure}

In galaxies, the occupation number of wave DM particles is huge and thus the system can be described by a classical wavefunction $\psi(t,\mathbf{r}) $. Such a system can be simulated in a cubic box of length $L$, and the dynamical equations are \cite{2017PhRvD..95d3541H}

\begin{equation}\label{eq:sch_1}
    i\hbar\frac{\partial\psi}{\partial t} = \bigg[- \frac{\hbar^2}{2m}\nabla^2 + m\Phi\bigg]\psi,
\end{equation}
\begin{equation}\label{eq:sch_2}
    \nabla^2\Phi = 4 \pi G M (|\psi|^2-\overline{|\psi|^2}),
\end{equation}
where $\Phi$ is the Newtonian gravitational potential, $m$ is the mass of the DM particle, and $M$ is the total DM mass in the box.
In our convention, the normalization condition is $\int_\mathrm{box}|\psi(t,\mathbf{r})|^2\mathrm{d}^3\mathbf{r} = 1$, and thus the spatial average $\overline{|\psi|^2} = V^{-1}$ with $V=L^3$.
In the simulation, we set $L = 10\,\mathrm{kpc}$, $M = 10^{11}M_\odot$, and mass of ultralight bosons, $m$, is given in the later text. The classical spectral algorithm was adopted to numerically evolve the system \citep{2017MNRAS.471.4559M,2021ApJ...910...29M}. The spatial resolution is $512^3$. The initial condition is set to be a collection of Gaussian wave packets. We first let them evolve long enough in time ($t_\mathrm{evo} = 5.11 \times 10^9\,\mathrm{yrs}$, with a relative large time step) to merger and form an isolated core. The density distributions around the core at the end of this stage are shown in Fig. \ref{fig:M11}. Next, in order to calculate the stretching effect given by Eq. (\ref{eq:dnu}) and to discuss the lensing issues, we adopt a much smaller time step ($\delta\mathrm{t} = 50\,\mathrm{yrs}$) in the followings.

In a lensing system, the typical angular distance between images is approximately the Einstein radius, which corresponds to $\sim$3\,kpc from the center in the lens plane. Therefore we take this length scale to calculate the result of Eq. (\ref{eq:dnu}). In the calculation, we only trace the potential energy change of each photon as it travels through the lens galactic halo but do not consider the deflection of the light path.

\begin{figure}
    \centering
    \includegraphics[width=0.95\linewidth]{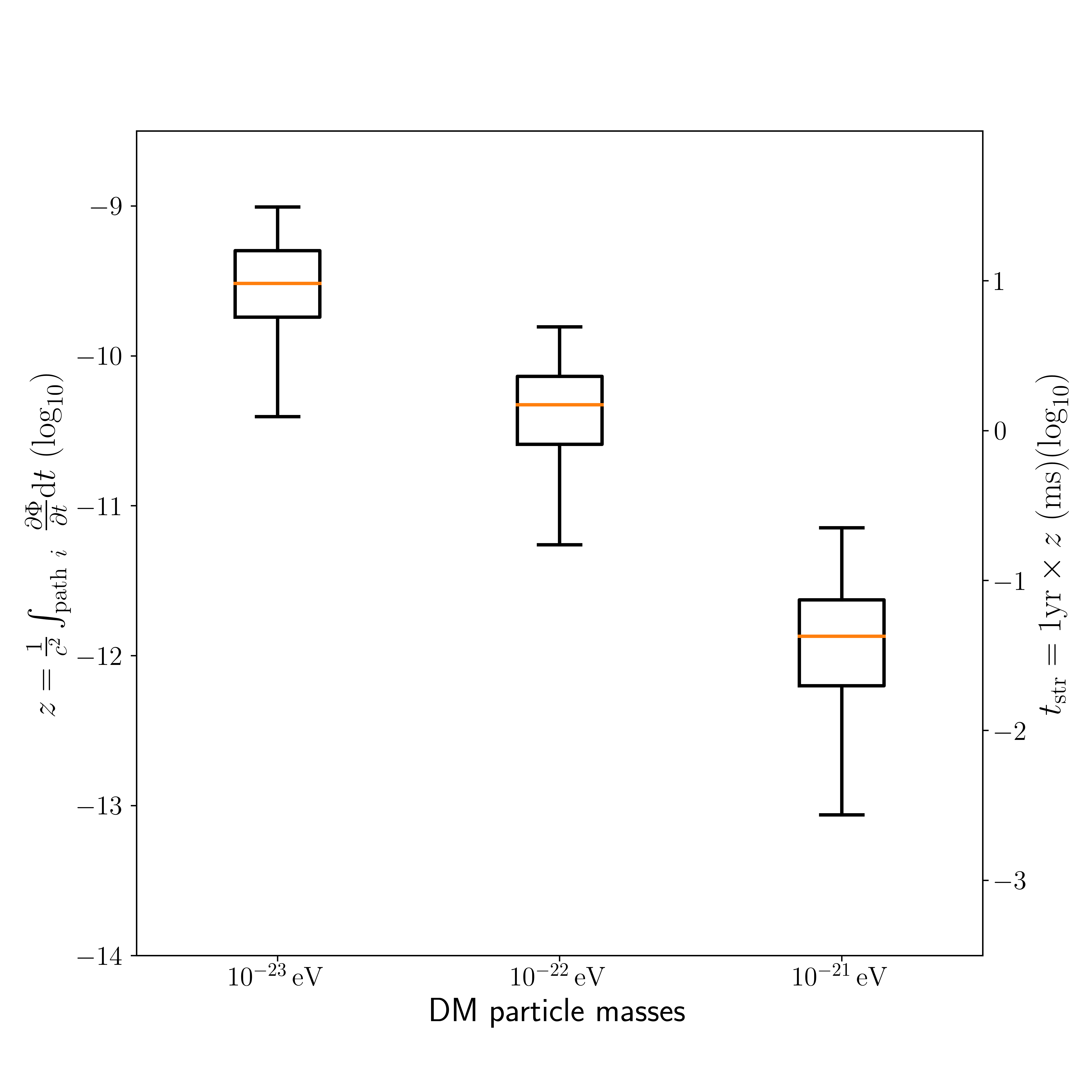}
    \caption{The magnitude of the stretching effect for different DM particle masses, with the right vertical axis corresponding to the stretch length for a one year sequence signal. The box diagram does not show outliers.}
    \label{fig:pic_4}
\end{figure}

\begin{figure}
    \centering
    \includegraphics[width=0.95\linewidth]{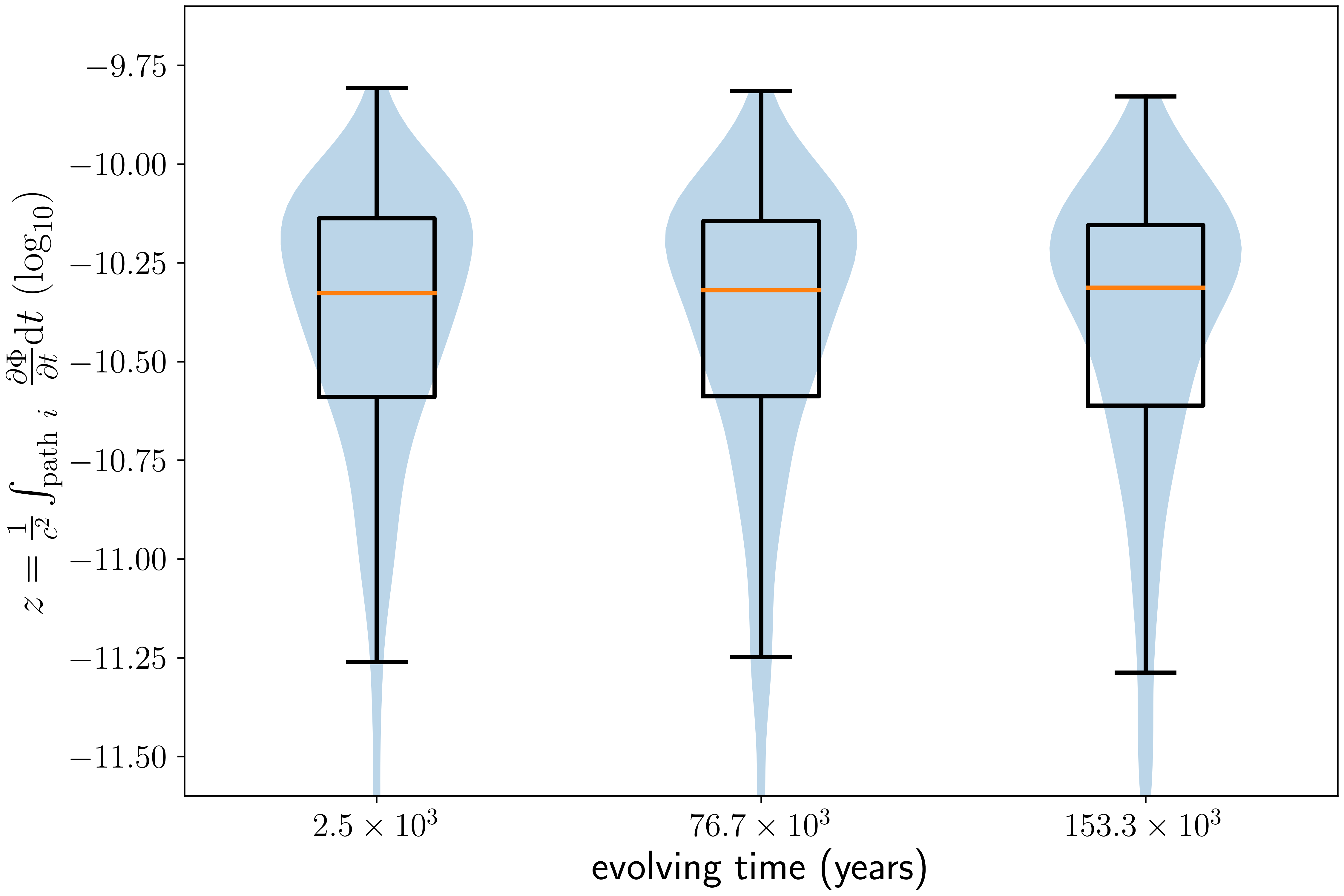}
    \caption{Results for $2.5 \times 10^3$, $76.7 \times 10^3$ and $153.3 \times 10^3$ years of evolution in the second evolutionary stage. The vertical coordinate is the calculation for Eq. (\ref{eq:dnu}), i.e. the value of the stretching effect. The other parameters are $m=10^{-22}\mathrm{eV},M=10^{11}M_\odot$ respectively.}
    \label{fig:m22}
\end{figure}

From the simulated DM halo made of $m\sim 10^{-22}~\mathrm{eV}$ ultralight bosons, we obtain that the magnitude of the stretching effect is $z\sim10^{-10}$. Moreover, we also perform simulations with $m\sim 10^{-21}\,\mathrm{eV}$ and $10^{-23}\,\mathrm{eV}$ taking into account, and derive corresponding magnitudes of the stretching effect. Results are presented in Fig.~\ref{fig:pic_4}. These results are in good agreement with theoretical predictions. That is, the halo made of lighter bosons exhibits stronger wave interference, and thus leads to larger variation of the gravitational potential with respect to time, ultimately increases the magnitude of the stretching effect. In addition, we also test the stability of simulations and subsequent estimations with different evolution time taking into consideration. Results are presented in Fig.~\ref{fig:m22} and suggest that our final estimations for the 
magnitude of the stretching effect induced by the wave DM halo is valid. However, as shown in Figs. \ref{fig:pic_4} and \ref{fig:m22}, this effect is extremely tiny, corresponding to several milliseconds stretching in one year for a sequence signal, and almost impossible to be measured with traditional astronomical observations. Luckily, the emergency of FRBs might be able to change the situation. 

\section{Observational probe}
FRBs are energetic radio transients with very short durations ($\sim$milliseconds) at cosmological distance~\citep{2007Sci...318..777L,2019ARA&A..57..417C,2023RvMP...95c5005Z,2017Natur.541...58C}. For currently available events, a considerable portion of them are repeating sources. Based on their extragalactic origin and very high event rate, FRBs have been widely proposed as important tools for studying cosmology and astrophysics. Especially, gravitationally lensed repeating FRBs have been put forward as powerful probes due to the extremely precision measurements of arrival times for the bursts. 
For a lensed repeating source, when we detect a number of signals from the weaker image, it is in principle that their corresponding signal could be reported in the brighter image. 

For a set of radio bursts, errors of their arrival time measurements mainly originate from the uncertainties of dispersion measure ($\mathcal{DM}$). Due to complex morphologies of bursts, different dedispersion methods, i.e. the best SNR or the best frequency alignment, might yield slightly different $\mathcal{DM}$ values and introduce an additional error. 
This data processing error changes with the shape of the burst etc. The state-of-art level of this uncertainty budget can be as small as $0.001~\mathrm{cm^{-3}~pc}$~\citep{2022arXiv220813677L}. Conservatively, we set $\Delta\mathcal{DM}\sim0.1~\mathrm{cm^{-3}~pc}$ as an optimistic case.
In conventional search pipelines, it is reasonable to set this error budget as $1.0~\mathrm{cm^{-3}~pc}$. These two values are taken into consideration for the sake of comparison in the following analysis.

We define two images in the lensing system as $a$ and $b$. Their actual arrival times are, 
\begin{equation}
\begin{aligned}
    t_a &= t_0 \times (1+z_a) + C, \\
    t_b &= t_0 \times (1+z_b),
\end{aligned}
\end{equation}
respectively. Here $t_0$ is the arrival time for a static lens (without the stretching effect), $z_a$ and $z_b$ are the magnitudes of the stretching effect for the two images, and $C$ is the time delay. We then take the difference between the two time series ($z_a \gg z_b)$
\begin{equation}
    \Delta t = t_a - t_b = t_0 \times (z_a - z_b) + C \sim t_0 \times z_a +C,
\end{equation}
and the likelihood function
\begin{equation}
    \begin{array}{l}
    \mathrm{ln} \  p(\Delta t|t_0,\sigma_a,\sigma_b,z_a,C) \\= 
    -\frac{1}{2} \sum_n\frac{(\Delta t_n - t_{0n} \times z_a - C)^2}{\sigma_a^2+\sigma_b^2} + \mathrm{ln}\bigg(2\pi(\sigma_a^2+\sigma_b^2)\bigg),
    \end{array}
\end{equation}
to estimate the magnitude of the stretching effect. For simplicity, we assume $\sigma_a=\sigma_b=\Delta \mathcal{DM}$. With the EMCEE method, we get the final result as shown in Tab. \ref{tab:1}.
Therein, $N$ is the total number of bursts reported in one year (uniform distribution), $\Delta \mathcal{DM}$ is the error for each signal (Gaussian distribution), and the time delay between two images is $\sim10$ days, 
stretching effect $z_a = 10^{-10}$.

\begin{table}
    \centering
    \begin{tabular}{|c|c|c|}
         \hline
         \multirow{2}*{$\Delta \mathcal{DM}~(\mathrm{cm^{-3}\,pc})$}
               &\multicolumn{2}{c|}{Number of signals} \\ \cline{2-3}
          & $N=100$ & $N=1000$ \\
         \hline
         1.0  & $60\%$ & $20\%$\\
       \hline
       0.1  & $6\%$ & $2\%$\\
        \hline
    \end{tabular}
    \caption{Relative error at $z = 10^{-10}$ with different lens FRBs parameters.}
    \label{tab:1}
\end{table}

Comparing the results shown in Tab. \ref{tab:1} with those in Fig.~\ref{fig:pic_4}, it is surprising and exciting that the stretching effect of a time-varying gravitational potential originating from the wave interference of a galactic DM halo composed of $m\sim10^{-22}~\mathrm{eV}$ ultralight bosons could be directly probed by monitoring the images of lensed repeating FRBs for $\sim1$ year.

\section{Conclusions and discussions}
Wave nature of a DM halo made of ultralight bosons directly induces a slow de Broglie time scale variation of the gravitational potential. In this paper, we first derive a formula for estimating the effect of a time-varying potential on photon propagation. Next, we carry out numerical simulations to quantify the magnitude of the variation of potential with time in this kind of wave DM halo and further estimate the stretching effect on a sequence signal caused by this time-varying potential. We obtain that, in a typical wave DM halo ($M\sim 10^{11}M_{\odot}$, $m\sim 10^{-22}~\mathrm{eV}$), the magnitude of the stretching effect is $z\sim10^{-10}$, corresponding to a $\sim\mathcal{O}(1)$ millisecond stretching of the signal in $\sim\mathcal{O}(1)$ year. This effect is terrible tiny and very challenging for traditional method to probe it. Moreover, we propose lensed repeating FRBs as a powerful tool to probe this effect. It is intriguing that the stretching effect induced by a DM halo composed of ultralight bosons with mass being $\leq10^{-22}~\mathrm{eV}$ could be directly detected by monitoring the images in a lensed FRB system for $\sim\mathcal{O}(1)$ year. Currently, FRB observations are rapidly growing, this method would probably achieve direct probe for the wave nature of galactic DM halos in the upcoming future.

It should be pointed out that there are many other effects that can also lead to the stretch of a sequence signal and mimic a time-varying potential. The first one is the moving lens effect of transverse shifts. The numerical magnitudes of these effects have been previously investigated and results suggested that these effects cause changes on the order of seconds~\citep{2018ApJ...866..101Z}. One possible approach is to reconstruct the transverse shift from observations of multiple images, considering that the transverse shift is an effect of the lens as a whole, but the effect of wave DM is not directly related to position. Therefore, this effect could be separated from observations.
Another factor that can affect this is the effect of Hubble flow, which is relatively small and causes changes on the order of milliseconds to hundreds of milliseconds. If the angular distance between the source and the lens is very small, this effect would be almost vanish.
Another factor that can affect this is the effect of Hubble flow, which is relatively small and causes changes on the order of milliseconds to hundreds of milliseconds~\citep{2021A&A...645A..44W}. If the angular distance between the source and the lens is very small, this effect would almost vanish.
In addition, plasma lensing could also be a non-negligible factor. Although the plasma lens modulates the arrival time of the signal, as long as the system does not change over time, it does not affect our results. On the other hand, the impact of plasma lensing is frequency-dependent, which can be clearly distinguished from our results.

\begin{acknowledgments}
This work was supported by the National Key Research and Development Program of China Grant Nos. 2023YFC2206702, 2021YFC2203001, the National Natural Science Foundation of China under Grants Nos. 12322301, 12222302, 12275021, 12021003, 12405050, and 11920101003, National SKA Program of China (2022SKA0130100), the science research grants from the China Manned Space Project with No. CMS-CSST-2021-B11, the Strategic Priority Research Program of the Chinese Academy of Sciences, Grant No. XDB23040100,  the Interdiscipline Research Funds of Beijing Normal University, and the Fundamental Research Funds for the Central Universities.
\end{acknowledgments}

\bibliography{sample631}

\end{document}